
\input harvmac
\newcount\figno
\figno=0
\def\fig#1#2#3{
\par\begingroup\parindent=0pt\leftskip=1cm\rightskip=1cm\parindent=0pt
\baselineskip=11pt
\global\advance\figno by 1
\midinsert
\epsfxsize=#3
\centerline{\epsfbox{#2}}
\vskip 12pt
{\bf Fig. \the\figno:} #1\par
\endinsert\endgroup\par
}
\def\figlabel#1{\xdef#1{\the\figno}}
\def\encadremath#1{\vbox{\hrule\hbox{\vrule\kern8pt\vbox{\kern8pt
\hbox{$\displaystyle #1$}\kern8pt}
\kern8pt\vrule}\hrule}}

\overfullrule=0pt

%
\def\tilde{\widetilde}
\def\bar{\overline}

\font\zfont = cmss10 

\def\bigone{\hbox{1\kern -.23em {\rm l}}}
\def\ZZ{\hbox{\zfont Z\kern-.4emZ}}

\Title{hep-th/9505186, IASSNS-HEP-95-36}
{\vbox{\centerline{ON $S$-DUALITY IN ABELIAN GAUGE THEORY}}}
\smallskip
\centerline{Edward Witten}
\smallskip
\centerline{\it School of Natural Sciences, Institute for Advanced Study}
\centerline{\it Olden Lane, Princeton, NJ 08540, USA}\bigskip
\baselineskip 18pt

\medskip

\noindent
$U(1)$ gauge theory on ${\bf R}^4$ is known to possess
an electric-magnetic duality symmetry that inverts the coupling constant
and extends to an action of $SL(2,{\bf Z})$.
In this paper, the duality is studied on a general four-manifold
and it is shown that the partition function is not a modular-invariant
function but transforms as a modular form.
This result plays an essential role in determining a new low-energy
interaction that arises when
$N=2$ supersymmetric Yang-Mills theory is formulated on a four-manifold;
the determination of this interaction gives a new test of the solution
of the model and would enter in computations of the Donaldson invariants
of four-manifolds with $b_2^+\leq 1$.   Certain other aspects of abelian
duality, relevant to matters such as the dependence of Donaldson
invariants on the second Stieffel-Whitney class, are also analyzed.

\Date{May, 1995}

\newsec{Introduction}

$S$-duality asserts that certain four-dimensional gauge theories
are invariant under modular transformations acting on
\eqn\taudef{\tau={\theta\over 2\pi}+{4\pi i\over g^2},}
with $\theta$ and $g$ being a theta angle and gauge coupling constant.
For the by now classical case of $N=4$ supersymmetric Yang-Mills theory,
this assertion was tested in \ref\vafa{C. Vafa and E. Witten, ``A Strong
Coupling Test Of $S$-Duality,'' Nucl. Phys. {\bf B431} (1994) 3.}
by actually computing
the partition function of the theory (with gauge group $SO(3)$ or
$SU(2)$) on certain four-manifolds $X$, in some cases with a topological
twist.  It was found that modular invariance did hold, in an appropriate
sense: (i) modular transformations in general exchange the gauge group
$SU(2)$ with the dual group $SO(3)$; (ii)
the partition function is not a modular-invariant function
but transforms as a modular form, with a weight proportional to the
Euler characteristic of the four-manifold.

The first point was anticipated by Montonen and Olive
\ref\olive{C. Montonen and D. Olive, ``Magnetic Monopoles As Gauge
Particles?'' Phys. Lett. {\bf 72B} (1977) 117.} but the second perhaps requires
comment.  That the partition function transforms as a modular form
rather than a modular function, with a modular weight as observed,
can be interpreted to mean that in coupling an $S$-dual theory to
gravity, to maintain $S$-duality, one requires certain non-minimal
$c$-number couplings that involve the background gravitational field only,
of the general form $\int_X \left(B(\tau)\tr R\wedge \tilde R +C(\tau)
\tr R\wedge R\right)$; here
$\tr R\wedge \tilde R$ and $\tr R\wedge R$ are
the densities whose integrals are proportional to the Euler characteristic
and the signature.  $B$ and $C$ are chosen so that $e^B$ and $e^C$ are
modular forms (of weights chosen to cancel the anomaly);
for $N=4$ supersymmetric Yang-Mills theory, one can
take $C=0$ (as the modular weight depends only on the Euler characteristic).

\nref\cardy{J. Cardy and E. Rabinovici, Nucl. Phys. ``Phase Structure Of
${\bf Z}_p$
Models In The Presence Of A Theta Parameter,'' {\bf B205} (1982) 1.}
\nref\ocardy{J. Cardy, ``Duality And The Theta Parameter In Abelian
Lattice Models,'' Nucl. Phys. {\bf B205} (1982) 17.}
\nref\owilczek{A. Shapere and F. Wilczek, ``Self-Dual Models With
Theta Terms,'' Nucl. Phys. {\bf B320} (1989) 669.}
\nref\buscher{T. H. Buscher, ``Path Integral Derivation Of Quantum Duality
In Nonlinear Sigma Models,'' Phys. Lett. {\bf 201B} (1988) 466.}
\nref\rab{A. Giveon, M. Porrati, and E. Rabinovici, ``Target-Space
Duality In String Theory,'' Phys. Rept. {\bf 244} (1994) 77.}
This effect is perhaps not really surprising,
but given our limited understanding of $S$-duality it appears difficult
to explain it {\it a priori} in, say,
the $N=4$ theory; similarly, one does not know how to predict
the coefficients of the anomaly in putatively $S$-dual theories in which
computations on four-manifolds
have not yet been performed, such as the $N=2$, $N_f=4$
theory with gauge group $SU(2)$
\ref\sw{N. Seiberg and E. Witten, ``Monopoles, Duality, And Chiral Symmetry
Breaking In $N=2$ Supersymmetric QCD,'' Nucl. Phys. {\bf B431} (1994) 484.}
\nref\ws{N. Seiberg and E. Witten,
``Electric-Magnetic Duality, Monopole Condensation, and Confinement In
$N=2$ Supersymmetric Yang-Mills Theory,'' Nucl. Phys. {\bf B426} (1994)
19.}.  The first goal of the present paper is to explore this
phenomenon in a much simpler example, namely free $U(1)$ gauge theory
without charges, where everything can be understood rather explicitly.
(The $S$-duality of this theory is an observation that goes back essentially
to \refs{\cardy - \owilczek}.)
In this case, the modular anomaly is, as will become clear, quite analogous
to the transformation law for the dilaton under $R\to 1/R$ symmetry
in two dimensions, as described in  \refs{\buscher,\rab}.

One would suppose that in $S$-dual theories that contain dynamical
gravity -- string theory is of course the only known candidate -- the
modular anomaly is somehow canceled.  Knowing how the anomaly works
in field theory with gravity as a spectator may ultimately be helpful for
understanding the case with dynamical gravity.

The shift from the group
to the dual group also has an analog for the free abelian theory.
In fact,
even self-dual lattices play a special role rather as in the theory
of chiral bosons in two dimensions.

The computation that we will perform is also relevant to various more
complicated problems in which this $U(1)$ theory is approximately embedded.
For instance, after computing the modular anomaly in section two,
we will use it in section three as a necessary ingredient in order to
compute a new effective interaction in $N=2$ supersymmetric gauge theory
with gauge group $SU(2)$ on a general four-manifold.
The unique and consistent determination of this interaction is
an interesting check on the framework of \ws.

In section four, we discuss some further details of how duality
works in the $N=2$ theory on a general four-manifold.
The results of sections three are needed for integration over the $u$-plane
to compute Donaldson invariants of four-manifolds with $b_2^+=1$, and the
results of section four are needed for a precise derivation of the relation
of  the Donaldson invariants to the monopole invariants of four-manifolds
described in \ref\ewitten{E. Witten, ``Monopoles And Four-Manifolds,''
Math. Res. Lett. {\bf 1} (1994) 769.}.  Some of these themes will be further
pursued elsewhere.

\newsec{Analysis Of The Abelian Theory}

In what follows, $X$ is a four-manifold and $b_i,\,i=0\dots 4$
will denote the $i^{th}$ Betti number,
that is the dimension of the space of harmonic $i$-forms on $X$.
As a harmonic two-form in four dimensions can be decomposed as a sum of
self-dual and anti-self-dual pieces, we can write $b_2=b_2^++b_2^-$
with $b_2^+$ and $b_2^-$ the dimensions of the spaces of self-dual
and anti-self-dual harmonic two-forms.  The Euler characteristic of
$X$ is $\chi=\sum_{i=0}^4(-1)^ib_i$ and the signature is $\sigma=b_2^+-b_2^-$.

We will say that a not necessarily
holomorphic function $F$ transforms as a modular form of weight $(u,v)$
for a finite index
subgroup $\Gamma$ of $SL(2,{\bf Z})$ if for
\eqn\immo{\left(\matrix{ a & b \cr c & d \cr}\right)\in {\Gamma}}
one has
\eqn\jimmo{F\left(
{a\tau+b\over c\tau+d}\right)=(c\tau+d)^u {(c\bar\tau+ d)}^vF(\tau).}

Our main result concerning the partition function $Z(\tau)$ of the
$U(1)$ Maxwell theory on a four-manifold is that
\eqn\piro{Z(-1/\tau)=\tau^u\bar \tau^v Z(\tau),}
where
$(u,v)=\left((\chi+\sigma)/4,(\chi-\sigma)/4\right)$.
Note that as $(\chi\pm\sigma)/2=1-b_1+b_2^{\pm}$ is in general integral
but not necessarily divisble by two, the weights in \piro\ are half-integral
in general.  Since also $Z(\tau+1)=Z(\tau)$ (or $Z(\tau+2)=Z(\tau)$ if
$X$ is not a spin manifold, as explained momentarily), \piro\ implies that
$Z$ is modular of weight $(u,v)$.

Now we proceed to the analysis.  We consider a $U(1)$ gauge field
$A_m$ (a connection on a line bundle $L$),
with field strength $F_{mn}=\partial_mA_n-\partial_nA_m$;
we also set $F_{mn}^{\pm}=
{1\over 2}\left(F_{mn}\pm {1\over 2}\epsilon_{mnpq}F^{pq}\right)$.
On a four-manifold $X$ of Euclidean signature, we take the classical
action to be
\eqn\refrow{\eqalign{I= & {1\over 8\pi}\int_X d^4x\sqrt g\left({4\pi\over
g^2}F_{mn}F^{mn}+{i\theta\over 2\pi}{1\over 2}
        \epsilon_{mnpq}F^{mn}F^{pq}\right)\cr
      =& {i\over 8\pi}\int_X d^4x\sqrt g\left(\bar\tau F^+_{mn}F^{+mn}
-\tau F^-_{mn}F^{-mn}\right),\cr}}
with $\tau$ defined in \taudef.

Before proceeding, let us determine the periodicity in $\theta$.
We permit $L$ to be an arbitrary line bundle, so the general
constraint on periods of $F$ is simply the Dirac quantization law;
thus, if $U$ is any closed two-surface in $X$, $\int_UF$ is an arbitrary
integer multiple of $2\pi$.  If $X$ is a spin manifold, then the smallest
possible non-zero value of $J=\int d^4x\sqrt g\epsilon^{mnpq}F_{mn}F_{pq}$
is obtained by taking $X=U\times V$, with $U$ and $V$ being two-spheres,
and picking the gauge field so that $\int_UF_{12}=\int_VF_{34}=2\pi$
(with other components vanishing).  Then one gets $J=8(2\pi)^2$,
so the $\theta$-dependent part of the action is $i\theta$.  This means
that the theory is invariant under $\theta\to \theta+2\pi$, that is
$\tau\to \tau+1$.  On the other hand, if $X$ is not a spin manifold,
one can find a line bundle on $X$ such that $J=4(2\pi)^2$, leading
to invariance only under $\theta\to\theta+4\pi$ or $\tau\to\tau+2$.
\foot{To state this in a less computational way, $J/16\pi^2$ measures
$c_1(L)^2$, which for a spin manifold is an even integer (as the intersection
form on the second cohomology group is even) while on a four-manifold
that is not spin it is subject to no divisibility conditions.}
We will discuss at the end of this section what modification of the
theory would be needed to get full $SL(2,{\bf Z})$ invariance, including
$\tau\to \tau+1$, when $X$ is not spin.

We will determine the modular weight of the partition
function in two ways: first we simply calculate the partition
function and see what modular weight it has; second we give an {\it a priori}
explanation by manipulation of the path integral.  Apart from being
more conceptual, the second derivation could be extended to determine
the modular transformation law of correlation functions.

\subsec{The Computation}

The partition function of the $U(1)$ theory is a product of several
factors.  There is a sum over the isomorphism class of the line bundle
$L$.  Because the gauge group is abelian, the $L$ dependence comes entirely
from the value of the classical action for a connection on $L$ that minimizes
it; the sum over $L$'s will give a generalized theta function.
For each $L$ we must integrate over $b_1$ zero modes and evaluate
a determinant for the non-zero modes.

In computing the determinant, one can ignore the $\theta$ term, which is
a topological invariant.  The rest of the kinetic energy is proportional
to $1/g^2\sim {\rm Im}\,\tau$.  The $\tau$ dependence of the regularized
determinant
is then roughly a factor
of ${\rm Im}\,\tau^{-1/2}$ for every non-zero eigenvalue of the kinetic
operator of $A$.  The zero modes do not give factors of ${\rm Im}\,\tau$;
they are tangent to the space of classical minima, which is a torus of
dimension $b_1(X)$ and has a volume independent of ${\rm Im}\,\tau$.

Let $B_k$ be the dimension of the space of $k$-forms on $X$; of course,
the $B_k$ are infinite, but we can make them finite with a lattice
regularization.  The total number of modes of $A$ is thus $B_1$.
However, we only want to count modes modulo gauge transformations.
As the gauge parameter is a zero-form, the number of modes of infinitesimal
gauge transformations is $B_0$, but one (the constant mode) acts trivially
on $A$, so the number of $A$ modes modulo gauge transformations is
$B_1-B_0+1$.  Removing also the zero modes of $A$, the number of non-zero modes
mod gauge transformations is $B_1-B_0+1-b_1$ (since $b_0=1$,
this is the same as
$B_1-B_0+b_0-b_1$).  The $\tau$ dependence of the determinant is
thus
\eqn\hurfo{{\rm Im}\,\tau ^{{1\over 2}(b_1-1)}\cdot {\rm Im}\,\tau^{{1\over 2}
(B_0-B_1)}.}

In a lattice regularization, one would eliminate the last,
cutoff-dependent factor by simply including in the definition of the
path integral a factor of ${\rm Im}\,\tau^{1/2}$ for every integration variable
(every one-simplex on the lattice), and a factor of ${\rm Im}\,\tau^{-1/2}$
for every generator of a gauge transformation (every vertex or
zero-simplex).  Since the factors required
are defined locally, this eliminates the cutoff-dependence of the theory
while preserving locality.  With this definition
of the coupling to gravity, the $\tau $ dependence of the determinant
is simply a factor
\eqn\zurfo{{\rm Im}\,\tau^{{1\over 2}(b_1-1)}.}

Alternatively, the numbers $B_k$ would be regularized in a Pauli-Villars
regularization by replacing them  with something like
\eqn\yurfo{\Tr \,e^{-\epsilon\nabla_k}}
with $\epsilon$ a small parameter and $\nabla_k$ the Laplacian
on $k$-forms.  The small $\epsilon$ behavior of \yurfo\ can be worked
out using the short time behavior of the heat kernel; the various terms
can be written as the integrals over $X$ of local densities.  As these
terms are {\it local}, dropping
them and dropping $B_0$ and $B_1$ in \hurfo\ simply
amounts to a specific choice of how the theory is coupled to gravity.
The choice that leads to \zurfo\ is the most
minimal one in that it corresponds, for instance, to the usual
prescription in which factors of $g$ in the determinant come only from
zero modes.  (Notice that, as $b_1-1$ cannot be computed as a local
integral, no choice of the coupling to gravity will remove the factor
obtained in \zurfo.)

Now we come to the sum over line bundles $L$.  The object
$F/2\pi$ is a de Rham representative of the first Chern class
$m=c_1(L)$.  We will think
of $m$ as a point in $H^2(X,{\bf R})$ that lies in the lattice $\Lambda$
consisting of points with integer periods.  For a given
$L$, the real part of the action is minimized for a connection
such that the two-form $F$ is harmonic.  For $F$ harmonic,
the dual two-form $\tilde F_{mn}={1\over 2}
\epsilon_{mnpq}F^{pq}$ is also harmonic.  In terms of $m$,
the map $F\to \tilde F$ is a linear map $m\to *m$ with $*^2=1$;
note that $*m$ takes values in $H^2(X,{\bf R})$, but not necessarily
in $\Lambda$.  If $(m,n)$ is the intersection pairing on $H^2$,
which is integral for lattice points $m,n$, then we have for $F$ harmonic
\eqn\wehave{\eqalign{ (m,m) & = {1\over 16\pi^2}\int d^4x\sqrt g\epsilon
^{mnpq}F_{mn}F_{pq} \cr
(m,*m) & = {1\over 8\pi ^2}\int d^4x \sqrt g F_{mn}F^{mn}.\cr}}
For $F$ harmonic, the action can therefore be written
\eqn\yoyo{I={4\pi^2\over g^2}(m,*m)+{i\theta\over 2}(m,m)
={i\pi\over 2}\left\{\tau\left((m,m)-(m,*m)\right)+\bar \tau
\left((m,m)+(m,*m)\right)\right\}.}
The sum over line bundles therefore gives
a lattice sum
\eqn\zoyo{F(q)=\sum_{m\in\Lambda}q^{{1\over 4}(-(m,m)+(m,*m))}
\,\bar q^{{1\over 4} ((m,m)+(m,*m))}}
with $q=e^{2\pi i \tau}$.

The function $F(q)$ is very similar to the generalized theta functions
that appeared in the work of Narain \ref\narain{K. S. Narain, ``New Heterotic
String Theories In Uncompactified Dimension $<10$,'' Phys. Lett. {\bf 169B}
(1986) 41; K. S. Narain, M. H. Sarmadi, and E. Witten, ``A Note On
Toroidal Compactification Of Heterotic String Theory,'' Nucl. Phys. {\bf B279}
(1987) 369.}
on toroidal compactification of two-dimensional conformal field theory.
It transforms as a modular function of weight $(b_2^-,b_2^+)$ (for
a subgroup of $SL(2,{\bf Z})$, in general).
$b_2^+$ and $b_2^-$
enter as the number of positive and negative eigenvalues of the intersection
form on $H^2$.  For instance, if $X$ is ${\bf CP}^2$ with its usual
complex orientation, so $b_2^+=1$, $b_2^-=0$, then $F$ reduces
to the complex conjugate of a standard theta function:
\eqn\oyo{F=\sum_{n\in{\bf Z}} \bar q^{n^2/2}.}
In particular, this $F$ is modular of weight $(0,1/2)$.  Similarly,
with opposite orientation, ${\bf CP}^2$ has $(b_2^+,b_2^-)=(0,1)$
and gives for $F$ a standard theta function of modular weight $(1/2,0)$.

Including also the determinant of the non-zero modes,
the Maxwell partition function up to a $\tau$-independent multiplicative
constant (which depends on the Riemannian metric of $X$)
is
\eqn\ubu{Z(\tau)=({\rm Im}\,\tau)^{{1\over 2}(b_1-1)}F(\tau).}
As
\eqn\hubu{{\rm Im}\,(-1/\tau)={1\over \tau\bar\tau}{\rm Im}(\tau),}
${\rm Im}\,(\tau)$ is modular of weight $(-1,-1)$.  So the Maxwell
partition function is modular of weight
\eqn\pubu{(u,v)={1\over 2}(1-b_1+b_2^-,1-b_1+b_2^+)={1\over 4}(\chi-\sigma,
\chi+\sigma).}
In particular, as in \vafa,  the modular weights
are linear in $\chi$ and $\sigma$.

\subsec{{\it A Priori} Explanation}

\def\cf{{\cal F}}
Now we proceed to a more {\it a priori} explanation of the above result,
along lines similar to a familiar discussion of $T$-duality
in two-dimensions \nref\rocek{M. Rocek and E. Verlinde, ``Duality, Quotients,
And Currents,'' Nucl. Phys. {\bf B373} (1992) 630.}
\refs{\buscher,\rocek}
(see also section 3.2 of
\ws\ where a version of this argument is given, on flat
${\bf R}^4$, including supersymmetry).

First we rewrite the Maxwell theory in terms of some additional
degrees of freedom.  We introduce a two-form field $G$.
We want to consider a theory which beyond
the usual Maxwell gauge invariance
$A\to A-d\epsilon$, $G\to G$, has invariance
under
\eqn\ubbu{\eqalign{A\to A + B \cr
                   G\to G + d B \cr}}
where $B$ is an arbitrary one-form or more generally a connection
on an arbitrary line bundle $M$.  Note that if $A$ is a connection on
a line bundle $L$, then $A+B$ should be interpreted as a connection
on $L\otimes M$.  Thus, while the ordinary Maxwell theory is well-defined
for any given $L$, its extension to include the $G$ field can only
have the invariance \ubbu\ if we sum over all $L$'s.
Invariance under \ubbu\ means that one can shift the periods of
$G$ -- that is the integrals of $G$ over closed two-dimensional
cycles $S\subset X$ -- by multiples of $2\pi$:
\eqn\bubbu{\int_S G \to \int_SG+2\pi n.}
Note that the usual Maxwell gauge invariance is a special case of \ubbu\
obtained by setting $B=-d\epsilon$.

There is an obvious way to achieve invariance under \ubbu: we
set $\cf= F- G$ and replace $F$ everywhere in the Maxwell Lagrangian
by $\cf$.  The resulting theory however is trivial and in particular
not equivalent to Maxwell theory.  To get something of interest, we
introduce a dual connection $V$ on a dual line bundle $\tilde L$,
with field strength $W_{mn}=\partial_mV_n-\partial_nV_m$.
Like the curvature on any line bundle, $W$ obeys a Dirac quantization law
\eqn\vubbu{\int_S W\in 2\pi {\bf Z}.}
This ensures that
\eqn\impo{{1\over 8\pi}\int_Xd^4x \sqrt g\epsilon^{mnpq}W_{mn}G_{pq}}
is invariant under \ubbu\ modulo $2\pi$.
We then take the Lagrangian to be
\eqn\wimpo{I={i\over 8\pi}\int_Xd^4x \sqrt g\left(\epsilon^{mnpq}W_{mn}G_{pq}
+\bar \tau \cf^+_{mn}\cf^{+mn}
-\tau \cf^-_{mn}\cf^{-mn}\right).}
The terms involving $\cf$ are manifestly invariant under
\ubbu\ (which acts trivially
on $V$; $V$ transforms nontrivially only under the dual gauge transformations
$V\to V-d\alpha$), and the same is true mod $2\pi i $ of the
first term by virtue of what has just been said.

We now proceed as follows.  After showing that \impo\ is equivalent
to the original Maxwell theory (provided that one sums over all $L$'s
in each case), we will show how to obtain by a different manipulation
a dual version.  We first perform both computations in a somewhat cavalier
manner and then more carefully study the quantum integration measure
to compute the modular weight.

For the first manipulation, the part of the path integral that involves
$V$ is
\eqn\uhu{\int DV\,\exp\left({i\over 8\pi}\int_Xd^4x \sqrt g
\epsilon^{mnpq}W_{mn}G_{pq}\right).}
The integral over $V$ consists of a discrete sum over dual line bundles
$\tilde L$ and for each $\tilde L$ a continuous integral over connections
on $\tilde L$.  The continuous part of the integral gives a delta
function setting $dG=0$.  The sum over $\tilde L$ then gives a further
delta function setting the periods of $G$ to be integral multiples of
$2\pi$.  But the exotic gauge invariance \ubbu\ permits one
(in a fashion that is unique up to an ordinary
gauge transformation) to set $G=0$ precisely if $dG=0$ and $G/2\pi$
has integral periods.  So after integrating over $V$, we gauge $G$ to
zero, reducing the extended gauge invariance to ordinary gauge invariance,
and reducing the Lagrangian \wimpo\
to the original Lagrangian \refrow\ of the abelian gauge theory.

The other way to analyze \wimpo\ begins by noting that the gauge invariance
\ubbu\ precisely enables one to fix a gauge with $A=0$.  In that
gauge, one can then integrate out $G$, giving a Lagrangian containing
$V$ only.  This Lagrangian inherits the dual gauge invariance
$V\to V-d\alpha$ of \wimpo, and is simply an abelian gauge theory
of the general type we started with in \refrow, but with a different
value of $\tau$.  The computation is very quick if one notes that in
terms of the the self-dual and anti-self-dual projections $W^\pm$
and $G^\pm$ of $W$ and $G$, the action is
\eqn\piloo{{i\over 4\pi}\int_Xd^4x\sqrt g\left(W^+G^+-W^-G^-\right)
+{i\over 8\pi}\int_Xd^4x\sqrt g\left(\bar \tau (G^+)^2-\tau(G^-)^2\right).}
Integrating out $G^\pm$ at the classical level, one gets simply
\eqn\jiloo{{i\over 8\pi}\int_Xd^4x\sqrt g\left(\left({-1\over\bar \tau}\right)
(W^+)^2-\left({-1\over\tau}\right)(W^-)^2\right).}
This is the original Lagrangian \refrow, but with $\tau\to -1/\tau$.

So far, we have integrated out various fields classically, without
taking account of certain determinants that will give the modular
anomaly.  To see the anomaly, we will now go over some of the
above steps in a more precise way.  Let $I_\tau(A)$ be the Maxwell
action \refrow\ with coupling parameter $\tau$.  Then we define
the partition function of the original theory to be
\eqn\orpar{Z(\tau)=({\rm Im}\,\tau)^{{1\over 2}(B_1-B_0)}
{1\over {\rm Vol}(G)}\int DA\,e^{-I_\tau(A)}.}
Here ${\rm Vol}(G)$ is the volume of the group of gauge transformations, and
$B_k$ is as before the dimension of the space of $k$-forms.
Thus, in \orpar\ we are simply including a factor of ${\rm Im}\,\tau^{1/2}$
for every integration variable and a factor of ${\rm Im}\,\tau^{-1/2}$
for every gauge generator, to cancel what would otherwise be the
cutoff-dependence of the path integral.
(We have already seen that the recipe in \orpar\ leads to a cutoff-independent
power of ${\rm Im}\,\tau$ in the final result for the partition function.)
Of course, the partition function can equally well be written in dual
variables
\eqn\porpor{Z(\tau)=({\rm Im}\,\tau)^{{1\over 2}(B_1-B_0)}
{1\over {\rm Vol}(\tilde G)}\int DV\,e^{-I_\tau(V)},}
with $\tilde G$ the dual gauge group.

The above derivation began with the fact that $Z(\tau)$ can alternatively
be defined by
\eqn\zorpar{Z(\tau)=({\rm Im}\,\tau)^{{1\over 2}(B_1-B_0)}
{1\over {\rm Vol}(K)\times {\rm Vol}(\tilde G)}\int DA\,DG\,DV\,
e^{-\hat I_\tau(A,G,V)}}
where $\hat I_\tau$ is the extended Lagrangian in \wimpo\ with coupling
parameter $\tau$, and $K$ is the extended gauge group.
Note that as the $V$-dependent part of $\hat I$ is independent of $\tau$,
one gets no factors of $\tau $ in integrating over $V$; the $V$ integral
gives a $\tau$-independent delta function by means of which the $G$ integral
can be done without generating any powers of ${\rm Im}\,\tau$.
That is why the same power of ${\rm Im}\,\tau$ appears
in \zorpar\ as in \orpar.

On the other hand, suppose that we evaluate \zorpar\ by gauging
$A$ to zero and then integrating over $G$.
One gets a factor of $\bar\tau^{-1/2}$ or $\tau^{-1/2}$ from
the integral over any mode of $G^+$ or $G^-$, so altogether
the $\tau$-dependent factor that one obtains from the $G$ integral
is $\tau^{-{1\over 2}B_2^-}\bar\tau^{-{1\over 2}B_2^+}$ with
now $B_2^\pm$ the dimensions of the spaces of self-dual and anti-self-dual
two-forms.  So eliminating $V$ and $G$ in \zorpar\ actually gives
(apart from a possible $\tau$-independent factor)
\eqn\jorpar{Z(\tau)= ({\rm Im}\,\tau)^{{1\over 2}(B_1-B_0)}
\tau^{-{1\over 2}B_2^-}\bar\tau^{-{1\over 2}B_2^+}{1\over {\rm Vol}(\tilde G)}
\int DV\,e^{-I_{-1/\tau}(V)}.}
Comparing this to \porpor, we get
\eqn\torpor{Z(\tau)=\tau^{-{1\over 2}(B_0-B_1+B_2^-)}\bar\tau^{-{1\over 2}
(B_0-B_1+B_2^+)}Z(-1/\tau).}
With $B_0-B_1+B_2^\pm = b_0-b_1+b_2^\pm=(\chi\pm\sigma)/2$,
this is equivalent to
\eqn\zorpor{Z(\tau)=\tau^{-{1\over 4}(\chi-\sigma)}\bar\tau^{-{1\over 4}
(\chi+\sigma)}Z(-1/\tau),}
so as we found earlier, $Z$ is modular
(for  $SL(2,{\bf Z})$ or a subgroup)
with weights $\left((\chi-\sigma)/4,(\chi+\sigma)/4\right)$.

\subsec{Full Modular Invariance?}

Finally, one might wonder how one must modify the construction if
one wants to obtain $SL(2,{\bf Z})$ covariance (and not just covariance
under a subgroup) even when $X$ is not a spin manifold.

In fact, we will embed this question in a somewhat broader question
of abelian gauge theory with gauge group $U(1)^n$ for arbitrary positive
integer $n$.  We introduce $n$ $U(1)$ gauge fields $A^I$, $I=1,\dots, n$,
with curvatures $F^I=dA^I$.   Picking a positive quadratic form
on the Lie algebra of $U(1)^n$ -- which we represent by a symmetric
positive definite matrix $d_{IJ}$ -- we take the Lagrangian to be
\eqn\ruff{\eqalign{
\, & {1\over 8\pi}\sum_{I,J} d_{IJ}\int_X d^4x\sqrt g\left({4\pi\over
g^2}F^I_{mn}F^{Jmn}+{i\theta\over 2\pi}{1\over 2}
        \epsilon_{mnpq}F^{Imn}F^{Jpq}\right)\cr
      =& {i\over 8\pi}\sum_{IJ}d_{IJ}
\int_X d^4x\sqrt g\left(\bar\tau F^{I+}_{mn}F^{J+mn}
-\tau F^{I-}_{mn}F^{J-mn}\right).\cr}}
In the space of $F^I$'s there is an integral structure given by the
fact that
for $S$ a closed two-surface in $X$, the quantities
\eqn\gruffu{\int_S{F^I\over 2\pi}}
are integers.  Thus, $d$ can be considered to define a quadratic form
on a certain lattice $\Gamma$.

To implement duality, we
extend the gauge-invariance as before,
adding two-form fields $G^I$ and replacing
$F^I$ by $\cf^I=F^I-G^I$.  Also, we
add dual gauge fields $V_I$, with curvatures $W_I=dV_I$.
We introduce an extended action
which is the obvious generalization of \wimpo:
\eqn\owimpo{I={i\over 8\pi}
\int_Xd^4x \sqrt g\left(\sum_I\epsilon^{mnpq}W_{Imn}G^I_{pq}
+\sum_{I,J}d_{IJ}\left(\bar \tau \cf^{I+}_{mn}\cf^{J+mn}
-\tau \cf^-_{Imn}\cf^{J-mn}\right)\right).}
Obvious generalizations of the previous manipulations show, on the
one hand, that after integrating out $V$ and $G$, the theory
defined by \owimpo\ is equivalent
to the theory defined by \ruff, and on the other hand, that after
picking the gauge $A=0$ and integrating out $G$, \owimpo\ can be
replaced by the dual Lagrangian
\eqn\gimpo{{1\over 8\pi}\sum_{I,J} d^{IJ}{i\over 8\pi}
\int_X d^4x\sqrt g\left(\left({-1\over \bar\tau}\right)W^{+}_{Imn}W_J^{+mn}
-\left({-1\over \tau}\right)W^{-}_{Imn}W_J^{-mn}\right).}
Here $d^{IJ}$ is the inverse matrix to $d_{IJ}$.

Now we can easily determine the conditions for modular covariance.
The transformation $\tau\to -1/\tau$ maps \ruff\ to \gimpo, so one has
covariance under this transformation if and only if the lattice $\Gamma$
with quadratic form $d$ is equivalent to the dual lattice with quadratic
form $d^{-1}$, that is if and only if the quadratic form
defined by $d$ is self-dual.
If this quadratic form is also {\it integral}, then one also has
invariance under $\tau\to \tau+2$ for any $X$, or $\tau\to \tau+1$ if
$X$ is spin, by an argument given earlier in our original discussion
of $U(1)$ gauge theory.  If one wants invariance under $\tau\to\tau+1$
for {\it any} $X$, not necessarily spin, then one must pick the quadratic
form to be {\it even}, not just integral.  This ensures that
\eqn\huvvo{{1\over 16\pi^2}\sum_{I,J}d_{IJ}\int_X d^4x\sqrt g
\epsilon^{mnpq}F^I_{mn}F^J_{pq} }
is divisible by two whether or not $X$ is spin, so that
the Lagrangian \ruff\ is invariant mod $2\pi i$ under
$\theta\to\theta+2\pi$, that is, under $\tau\to\tau+1$.
It is curious that even integral lattices are thus related to modularity
in four dimensions as they are in two dimensions.

\newsec{A New Effective Interaction On The $u$ Plane}

In \ws, the low energy dynamics of the pure $N=2$ supersymmetric
gauge theory with gauge group $SU(2)$ was determined.  A familiarity
with that discussion will be assumed here, but a few points will be repeated
to fix notation.  The basic
order parameter in \ws\ was $u=\Tr\, \phi^2$ ($\phi$ being the scalar field
related by supersymmetry to the gauge field); the variables
in the low energy theory were $u$ and its superpartners.  In
the low energy theory, it is natural to count dimensions in such
a way that $u$ has dimension zero and its fermionic partners dimension
one-half.  With that way of counting, as long as one is on flat
${\bf R}^4$, the supersymmetric interactions of lowest dimension
have dimension two.  All interactions of dimension two were determined
in \ws\ in terms of functions associated with the family of elliptic curves
\eqn\uppu{y^2=(x^2-1)(x-u).}

If one works on a curved four-manifold, local interactions of dimension
zero -- involving couplings of $u$ to polynomials in the Riemann tensor
and its derivatives -- become possible.  In the physical theory on
a four-manifold, a rather complicated structure might be possible
in general, but in the twisted topological theory the only
possible such interactions are proportional to
$\int_X f(u)\,\,\tr R\wedge \tilde R$ or $\int_X g(u)\,\, \tr R\wedge R$
where $\tr R\wedge \tilde R$ and $\tr R\wedge R$ are the densities
whose integrals are proportional to $\chi$ and $\sigma$, respectively,
and the functions $f$ and $g$ are holomorphic.\foot{
BRST-invariant configurations have $u$ constant; for constant $u$ a function
$\int T(u) {\cal O} \sqrt g d^4x$, with ${\cal O}$ a locally constructed
function of the metric, is a topological invariant if and only if
${\cal O}$ is related to the Euler characteristic or the signature or is an
irrelevant total derivative.
Moreover, BRST invariance holds if and only if $T$ is holomorphic.}
Such interactions
produce in the path integral measure a factor which if $u$ is constant
simply takes the form
\eqn\ippopo{\exp\left(b(u)\chi +c(u)\sigma\right).}
In the present section, we will determine the functions $b$ and $c$.
These interactions are closely related to similar interactions
that appear in the untwisted physical
theory on a four-manifold, but the precise
relation will not be analyzed here.

The reasons for performing this computation and presenting the result
here are as follows:

(1) The computation turns out to involve the result of the last
section in an interesting way.

(2) The ability to get a unique and consistent result for $b$ and $c$
provides an interesting new check on the formalism of \ws.

(3) The result is needed for integrating over the $u$ plane in Donaldson
theory (in order to obtain formulas for Donaldson invariants of
four-manifolds of $b_2^+\leq 1$), though this application will not be
developed in the present paper.

\subsec{Asymptotic Behavior}

The obvious way to proceed (temporarily overlooking a subtlety
that will presently appear) is as follows.  Singularities and zeroes
in the holomorphic function \ippopo\
at finite places on the $u$ plane can only occur at $u=\pm 1$,
where extra massless particles (monopoles and dyons) appear,
giving rise to singularities in various physical quantities, including
those computed in \ws.  (Note that either a zero or a pole
in the function \ippopo\ is associated
with a singularity in $b$ or $c$ and so is possible only at $u=1,-1$, or
$\infty$.)  The behavior of the unknown function can be obtained for
$u\to \infty$ using asymptotic freedom and weak coupling, and for
$u\to \pm 1$ using a knowledge of which particles are becoming massless
at those points.
It would appear that knowledge of the behavior at infinity and at the
singularities would determine the sought-for holomorphic function
uniquely except for the possibility
of adding constants to $b$ or $c$.  Let us see how this program works
out.

First we consider the behavior for large $u$.  Let us recall that
the $N=2$ theory has an anomalous $U(1)_R$ symmetry which in
a field of instanton number $k$, on a four-manifold of given
$\chi$ and $\sigma$, has an anomaly
\eqn\huggu{\Delta R = 8k-{3\over 2}(\chi+\sigma).}
(That is, the operators with non-zero expectation value have $R$-charge
equal to $\Delta R$.)  The term in this formula of interest here is the part
involving the coupling to gravity, namely
\eqn\ippu{-{3\over 2}(\chi+\sigma).}
This has the following interpretation: the index theorem is such
that for each generator of the gauge group there is an anomaly
$-(\chi+\sigma)/2$; an extra  factor of three arises because $SU(2)$ is
three-dimensional.

For large $u$, $SU(2)$ is spontaneously broken to $U(1)$.  The zero modes
of the fermion fields in the $u$ multiplet carry the anomaly
$-(\chi+\sigma)/2$ of the one-dimensional
unbroken part of the group.  The remaining
anomaly $-(\chi+\sigma)$ must be manifested in an interaction
proportional to $\chi$ and $\sigma$ obtained by integrating out the
massive $SU(2)$ partners of the light fields.
This interaction must have no derivatives (or it would not
lead to violation of the symmetry when $u$ is constant)
so it is of the form \ippopo.  We can therefore determine
the large $u$ behavior of \ippopo: as $u$ has $R$-charge four, we need
\eqn\binvo{e^{b\chi+c\sigma}\sim u^{(\chi+\sigma)/4}~{\rm for}~u\to\infty.}

The behavior near $u=\pm 1$ can be determined similarly.  The effective
theory near $u=1$ has an accidental low energy $R$ symmetry.
The anomaly in this symmetry is
\eqn\immob{-{1\over 2}(\chi+\sigma)+{c_1(L)^2\over 4}-{\sigma\over 4}.}
Here
$-(\chi+\sigma)/2$ is the contribution from the $u$-multiplet and is
carried by the zero modes related to $u$ by supersymmetry,
and $ c_1(L)^2/4-\sigma/4$
is the contribution of the monopoles that become massless
at $u=1$.  The gravitational part of the monopole contribution is
$-\sigma/4$ and must show up in a singular
behavior of the interaction $e^{b\chi+c\sigma}$
obtained by integrating out the light monopoles.
(The $c_1(L)^2/4$ appears in the behavior of the effective gauge couplings
near $u=1$.)  As
$u-1$ has charge two under the $R$ symmetry near $u=1$, the singular
behavior is
\eqn\ubinvo{e^{b\chi+c\sigma}\sim (u-1)^{\sigma/8} ~{\rm for}~u\to 1.}
By a similar argument, the singular behavior near $u=-1$, obtained
by integrating out light dyons that again violate the effective $R$-symmetry
by $-\sigma/4$, must be
\eqn\nubinvo{e^{b\chi+c\sigma}\sim (u+1)^{\sigma/8}~{\rm for}~u\to -1.}

We are left, then, looking for a holomorphic function with the
singularities and zeroes
given in \binvo, \ubinvo, and \nubinvo.
For the terms involving $\sigma$ there is an evident and unique
formula:
\eqn\hinbo{e^{c\sigma}=(u^2-1)^{\sigma/8}.}
(When this is multiplied by the partition function
of the massless photon, allowing for the fact that as explained in
section four the ``dual line bundle'' is really a ${\rm Spin}^c$ structure,
the branch cuts that are present
if $\sigma$ is not divisible by eight disappear.)
But for the terms involving $\chi$ we face a quandary:
no holomorphic function $e^{b\chi}$ can have the asymptotic
behavior given by \binvo\ and the lack of singularities or zeroes
indicated in the subsequent formulas.

\subsec{Modular Anomaly}

To unravel this puzzle, we have to recall in somewhat more depth the
structure found in \ws.

There is no completely canonical way to describe the low energy
effective action of this theory.  Over the $u$-plane punctured at
$u=1,-1$, there is a flat $SL(2,{\bf Z})$ bundle $E$, the fiber at a point
$u$ being the first cohomology group of the elliptic curve described in
equation
\uppu.  Every local trivialization of $E$ gives a way of writing
an effective action; different trivializations give different formulas
related by $SL(2,{\bf Z})$ transformations.

Periods of a certain differential form (described precisely in \ws, section 5)
give a certain holomorphic section $(a_D,a)$ of $u$.  Either $a_D$ or $a$
(or any integer linear combination) can be taken as the basic variable
in the low energy description; they are related by supersymmetry
to gauge fields $A_D$ and $A$ respectively.
$A_D$ and $A$ are dual gauge fields
related by the duality transformation of the previous
section.  The transformation from $a$ to $a_D$ as the basic variable
is a special case of an $SL(2,{\bf Z})$ transformation, achieved by
the matrix
\eqn\hmat{\left(\matrix{ 0 & 1 \cr -1 & 0 \cr}\right):\left(\matrix{
a_D \cr a \cr}\right)\to\left(\matrix{
a \cr -a_D \cr}\right).}
The gauge coupling constant is expressed in terms of $a_D$ and $a$
by
\eqn\jmat{\tau = {da_D\over da}.}
It therefore transforms under \hmat\ by
\eqn\kmat{\tau\to -{1\over \tau}.}
As we have seen in the last section, this is how the coupling transforms
when one changes from the description by $A$ to the description by $A_D$.

Now let us discuss how the {\it quantum} low energy theory transforms
under duality.   In brief, we will find an anomaly that will precisely
cancel the difficulty that we had above in determining the function
$e^{b\chi}$.  Since the dimension two part of the
classical action is duality-invariant (this being
part of the construction in \ws) and since the operators (such as $u$)
whose correlations one wishes to compute are also completely duality
invariant, the question of invariance of the quantum theory under duality
amounts
to the question of how the quantum integration measure transforms under duality
and how the function $e^{b\chi}$ transforms.
We really only need to consider the special modular transformation
$\tau\to -1/\tau$, since $SL(2,{\bf Z})$ is generated by this transformation
together with $\tau\to\tau+1$, which acts trivially on the integration
variables and so produces no anomaly in the measure.  To understand
what happens under $\tau\to -1/\tau$, one must examine three things:

(1) The integration measure of $a$ and $\bar a$ (or $a_D$ and
$\bar a_D$).

(2) The fermion integration measure.

(3) The integration measure of the gauge fields.

We will work out the behavior of the three measures under duality as follows.

(1)  Because the kinetic energy of $a$ and $\bar a$ is proportional to ${\rm
Im}
\,\tau$,
the integration measure for those fields has the form
\eqn\thefo{{\rm Im}\,\tau \,\,da\,d\bar a.}
(This is true for every mode of $a,\bar a$, and so most importantly
it is true for the zero modes.)  Setting $\tau_D=-1/\tau$, one has
\eqn\offo{{\rm Im}\,\tau  \,da\,d\bar a={\rm Im}\,\tau_D\,\,da_D\,d\bar a_D}
so the integration measure for these fields is completely duality invariant.

(2) In contrast to the scalars, which have just been seen to have
no modular anomaly, the fermions do have such an anomaly.  The reason for this
is that  in \ws, duality acts on the fermions via a chiral transformation,
which has an anomaly on a curved four-manifold.

The details can be worked out as follows.  Like that of the scalars,
the fermion kinetic energy is proportional to ${\rm Im}\,\tau$,
so the integration measure for any normalized fermi mode $\beta$ is
\eqn\defro{d(\sqrt{{\rm Im}\,\tau}\,\beta)={d\beta\over\sqrt{{\rm Im}\,\tau}}.}
The fermi modes can be divided into modes of $R=1$ and $R=-1$,
which we will generically  call $\alpha$ and $\bar\alpha$, respectively.
\foot{In the twisted theory, $\alpha$ is a one-form and $\bar \alpha$
a linear combination of a zero-form and a self-dual two-form.}
It will be evident from the structure of what follows that only
zero modes need to be considered, the other fermion modes canceling
in pairs.  The duality transformation in \ws\ was such that under
$\tau\to -1/\tau$, $\alpha$ transforms to $\alpha_D=\tau\alpha$ and
$\bar\alpha$ transforms to $\bar\alpha_D=\bar\tau\,\bar\alpha$.
Hence
\eqn\noteth{\eqalign{{d\alpha\over \sqrt {{\rm Im}\,\tau}}
&=\sqrt{{\tau\over \bar \tau}}{d\alpha_D\over \sqrt{{\rm Im}\,\tau_D}} \cr
{d\bar \alpha\over \sqrt {{\rm Im}\,\tau}}
&=\sqrt{{\bar \tau\over \tau}}{d\bar \alpha_D\over \sqrt{{\rm
Im}\,\tau_D}}\cr}}
So if $d\mu^F$ is the fermion measure for $\alpha,\bar\alpha$
and $d\mu^F_D$ is the measure for $\alpha_D,\bar\alpha_D$, then one
has
\eqn\hoteth{d\mu^F=\left({\tau\over \bar \tau}\right)^{-(\chi+\sigma)/4}
d\mu^F_D,}
using the fact that the number of $\alpha$ zero modes minus the number
of $\bar\alpha$ zero modes is $-(\chi+\sigma)/2$.

(3) The transformation law for the gauge measure $d\mu^G$ is the most
subtle of the three and
can be read off from \zorpor:
\eqn\boteth{d\mu^G=\tau^{-{1\over 4}(\chi-\sigma)}\bar\tau^{-{1\over 4}(\chi
+\sigma)}d\mu^G_D.}
Actually, the formulation in \boteth\ is imprecise as $d\mu^G$ and
$d\mu^G_D$ are defined in different spaces.  This statement simply
means that the integral of $\mu^G$, that is the partition function,
differs from the integral of $\mu^G_D$ by the stated factor.

Multiplying the factors found in the above equations,
the relation between the measure
$d\mu$ of the theory using $a$ as the basic variable and the measure
$d\mu_D$ of the theory using $a_D$ is
\eqn\utoteth{d\mu=\tau^{-\chi/2}d\mu_D.}
The fact that the $\bar\tau$ dependence cancels here is crucial
in making it possible to cancel this anomaly by the holomorphic term
considered below.

\subsec{Final Determination}

It is now clear that $e^{b\chi}$ should not be a function of $u$
but rather should transform under duality as a holomorphic modular form
of weight $-\chi/2$.  The fact that the weight depends on $\chi$ and
not on $\sigma$ is the reason that above we found a straightforward
determination of the function $e^{c\sigma}$ but not a straightforward
determination of $e^{b\chi}$.  With the modular anomaly understood,
it is now easy to find the ``function'' $e^{b\chi}$:
\eqn\unco{e^{b\chi}=\left((u^2-1){d \tau\over du}\right)^{\chi/4}
.}

The first point of this formula is that if likewise
\eqn\bunco{e^{b_D\chi}=
\left((u^2-1){d \tau_D\over du}\right)^{\chi/4}
}
with $\tau_D=-1/\tau$, then
\eqn\hunco{e^{b\chi}=\tau^{\chi/2}e^{b_D\chi}}
so including in the theory a factor of
$e^{b\chi}$ cancels the modular anomaly.  Also, for $u\to \infty$,
$d\tau/d u \sim 1/u$ according to \ws, so \unco\ has the large $u$
behavior required in \binvo.

We still need to check \ubinvo, but first we must interpret it correctly.
\ubinvo\ was based on a computation near $u=1$, where the good
description is in terms of $a_D$.  (For $u$ away from the punctures
at $u=\pm 1$ and the singularity at $\infty$,
one can use either $a$ or $a_D$.)  Therefore, \ubinvo\
must be interpreted as a condition on the behavior of $e^{b_D\chi}$
near $u=1$, and this condition is obeyed since according to \ws\ one has
$d\tau_D/du\sim 1/(u-1)$ near $u=1$, ensuring that
$e^{b_D\chi}$ has no singularity at $u=1$.
Likewise, after transforming to the appropriate local description,
one finds the desired absence of singularity near $u=-1$.
(The local description near
$u=-1$ is \ws\ a third one using $a+a_D$ as the basic variable.)

The only point that remains is that the function $d\tau/du$ has neither
zeroes nor poles for finite $u$
away from $u=\pm 1$.  This is so because the family of curves
\uppu\ is the modular curve of the subgroup $\Gamma(2)$ of $SL(2,{\bf Z})$
(consisting of matrices congruent to one modulo two), and there are
no orbifold points in the moduli space.

Given that our candidate for $e^{b\chi}$ has the correct behavior
near $u=1,-1,$ and
$\infty$ and has no unwanted zeroes or poles, it must be the correct
answer (up to a multiplicative constant), since any other candidate
would be obtained by multiplying by an ordinary meromorphic
function which if not
constant would have unwanted zeroes and poles somewhere.

In sum, in any computation that involves
an integration on the $u$-plane, the more obvious factors in the path
integral must be supplemented by a new interaction that gives an
additional factor
\eqn\addfac{e^{b\chi+c\sigma}=
\left((u^2-1){d\tau\over du}\right)^{\chi/4}(u^2-1)^{\sigma/8}.}
This factor will enter in computations of Donaldson invariants
for $b_2^+\leq 1$.

\newsec{Abelian Duality Embedded In $SU(2)$ And $SO(3)$}

In this section, I will explain a few subtleties about the duality
transformation which -- in $N=2$ super Yang-Mills theory with gauge group
$SU(2)$ or $SO(3)$ --
relates the description in variables appropriate at large $u$
to the description valid near $u=1$. This will enable us to explain
some assertions made in \ewitten\ and will serve
as background for a further discussion which
will appear elsewhere.

First we must discuss how the $U(1)$ duality considered in section two
is embedded in $SU(2)$ or $SO(3)$.
To begin with, we assume that the gauge group is $SU(2)$
with rank two gauge bundle $F$.  At a generic point of the $u$ plane,
$SU(2)$ is broken down to $U(1)$, and $F$
splits as $F=T\oplus T^{-1}$ with $T$ a line
bundle.  The gauge field at low energies then
reduces to a $U(1)$ gauge field $C$, which one can think of as a connection
on $T$.

However, we want to be free to consider the case that the gauge group is
actually $SO(3)$, with a rank three gauge bundle $E$ which may have
$w_2(E)\not= 0$.  With the symmetry broken
to $U(1)$, $E$ decomposes at low energies as ${\cal O}\oplus L\oplus L^{-1}$,
with ${\cal O}$ a trivial bundle and $L$ a line bundle.  The connection reduces
at low energies to a $U(1)$ connection $A$ on $L$.
If $w_2(E)=0$, then the gauge group can be lifted to
$SU(2)$ and the rank two bundle $F$ exists; in that case $L=T^{\otimes 2}$
and $A=2C$.

Near $u=\pm 1$, one has instead a description using a dual gauge field
$V$ and dual ``line bundle'' $\tilde L$.
\foot{Note that $\tilde L$ was simply called $L$ in \ewitten\ --
the tilde was deleted as the ``original'' line bundle $L$ never
entered explicitly in that paper.}

There are two goals in this section:

(1) To explain how the description near $u=1$ (or $-1$) depends on
$w_2(E)$.

(2) To explain by what mechanism it turns out that the ``line bundle''
in the dual description near $u=1$ is really not a line bundle but
a ${\rm Spin}^c$ structure.

With the first goal in mind, it is clearly suitable to choose variables
adapted to the possibility that $w_2(E)\not=0$, so we will take
the basic variable to be $A$ rather than $C$.
Setting $F=dA$, let us now describe the Dirac quantization condition on
$F$ for general $w_2$.  For simplicity in exposition, we suppose that there
is no torsion in the cohomology of $X$ so that we can pick a basis of
two-dimensional closed surfaces $U_\alpha$ giving a basis of $H_2(X,{\bf Z})$.
Then $w_2(E)$ can be described by the conditions
\eqn\loppo{(w_2(E),U_\alpha)=c_\alpha}
with each $c_\alpha=0$ or 1.

The Dirac quantization condition asserts that
\eqn\dico{\int_{U_\alpha}F = 2\pi(2k_\alpha+c_\alpha),}
with $k_\alpha\in {\bf Z}$.  The idea behind this formula is that
if $C=A/2$, then Dirac quantization for $C$ fails precisely for those
$U_\alpha$ for which $c_\alpha=1$.

\subsec{Dependence On $w_2$}

We will now determine how the dual description near $u=1$
(or $-1$) depends on $w_2(E)$.
As in \ws, we take the starting Lagrangian for $A$ to be
\eqn\hujjo{I={i\over 16\pi}\int_Xd^4x\sqrt g\left(\bar\tau (F^+)^2
 -\tau (F^-)^2\right).}
(One can think of this as \ruff\ adapted to the root lattice of $SU(2)$,
which is generated by a point of length squared two; that is, \ruff\
turns into \hujjo\ if one sets $d=2$ and renames the variable called
$A$ in \ruff\ as $C=A/2$.  That rescaling also turns the standard Dirac
condition assumed in discussing \ruff\ to the special case $c_\alpha=0$ of
\dico.)

Now we introduce a dual $U(1)$ gauge field $V$, which is a connection
on a dual line bundle $\tilde L$.
We take the curvature $W=dV$
to obey standard Dirac quantization, that is $\int_{U_\alpha}W$ is an arbitrary
integer multiple of $2\pi$ without any refinement such as that in
\dico.  We also introduce a two-form $G$ and the extended gauge invariance
\eqn\uncub{\eqalign{A\to & A+2B \cr
                    G\to & G+2dB\cr}}
with $B$ a connection on an arbitrary line bundle $N$.  The reason for the
factor of two in \uncub\ is to preserve
the structure of \dico; that is, with this factor of two in the gauge
transformation law, $\int_{U_\alpha}F$ is gauge
invariant modulo $4\pi$.  Rather as in section two, we introduce
the gauge invariant object $\cf=F-G$ and the extended Lagrangian
\eqn\uxxxx{I'={i\over 16\pi}\int_Xd^4x\sqrt g \epsilon^{mnpq}W_{mn}G_{pq}
+{i\over 16\pi}\int_Xd^4x\sqrt g\left(\bar\tau (\cf^+)^2
 -\tau (\cf^-)^2\right)}
This has been chosen so that the theory defined by $I'$ is equivalent
to the theory defined by $I$.  To prove this, as in section two,
one integrates over $V$, obtaining a delta function that sets $G$ to
zero up to a gauge transformation; one uses the fact that the periods
of $G$ are gauge-invariant modulo $4\pi$.

To obtain a dual description, the first step in section two was to
use the extended gauge invariance to set $A=0$.
Now we cannot do that (unless $w_2(E)=0$) because of the factor of
two in \uncub.  The best that we can do is to select a fixed set of
line bundles $Q_\alpha$, with connections $\theta_\alpha$ and curvatures
$g_\alpha$,
such that
\eqn\oppop{\int_{U_\alpha}g_\beta=2\pi \delta_{\alpha
\beta},}
and use the extended gauge invariance to set $A=\sum_\alpha c_\alpha
\theta_\alpha$.
Then, by shifting $G$ by $G\to G+\sum_\alpha c_\alpha g_\alpha$,
the extended Lagrangian
turns into
\eqn\topop{I'=
{i\over 16\pi}\int_Xd^4x\sqrt g \epsilon^{mnpq}W_{mn}(G_{pq}-c_\alpha
g_{\alpha\,pq})
+{i\over 16\pi}\int_Xd^4x\sqrt g\left(\bar\tau (G^+)^2
 -\tau (G^-)^2\right).}
Integrating out $G$ now gives
\eqn\lotop{I''=
{i\over 16\pi}\int_Xd^4x\sqrt g\left(\left({-1\over \bar\tau}\right) (W^+)^2
 -\left({-1\over \tau}\right) (W^-)^2\right)
-{i\over 16\pi}\sum_\alpha c_\alpha\int_Xd^4x\sqrt g \epsilon^{mnpq}W_{mn}
g_{\alpha\,pq}.}

The $c_\alpha$ thus appear only in the last term, which is a topological
invariant, and moreover is always an integral multiple of $\pi i$
so that its exponential is $\pm 1$.
In fact, the exponential of the last term is
\eqn\ibbobb{ (-1)^{(c_1(\tilde L),w_2(E))}.}

This factor determines the dependence of the dual Lagrangian on $w_2(E)$.
It must be compared to the factor called $(-1)^{x'\cdot z}$
in eqn. (2.17) of \ewitten, which was claimed to determine
the dependence of the
Donaldson invariants on $w_2(E)$.
In \ewitten, $z$
was defined as $w_2(E)$, and the definition of $x'$ was such
that if $w_2(X)=0$, then $x'=-c_1(\tilde L)$.
\foot{In fact, $x$ was defined in \ewitten\ by $x=-2c_1(\tilde L)$,
and $x'$ by $2x'=x+w_2(X)$, so if $w_2(X)=0$, $x'=-c_1(\tilde L)$
and $(-1)^{x'\cdot z}$ coincides with the formula in \ibbobb.}
So for spin manifolds,
$(-1)^{x'\cdot z}$ coincides with \ibbobb.

To generalize this discussion to manifolds
that are not spin, there are additional subtleties to which we now turn.

\subsec{Appearance Of  ${\rm Spin}^c$ Structures}

The low energy theory near
$u=1$ has light magnetic monopoles.
In the topologically twisted version which leads to the
considerations in \ewitten, the ``monopoles,'' if $w_2(X)=0$,
are sections of $S_+\otimes \tilde L$,
where $S_+$ is the  positive chirality spin bundle and $\tilde L$ is
the dual line bundle.  If $w_2(X)\not= 0$, then $S_+$ does not exist
and it is clear, pragmatically, that the monopoles must be sections of
a ${\rm Spin}^c$ bundle that we can informally write as $S_+\otimes \tilde L$
but is no longer really defined as a tensor product.  (A precise
description of this situation is that $\tilde L^{\otimes 2}$ is
an ordinary line bundle such that $x=-c_1(\tilde L^{\otimes 2})$ is
congruent modulo two to $w_2(X)$; and $S_+\otimes \tilde L$ is a
${\rm Spin}^c$ bundle with an isomorphism $\wedge^2(S_+\otimes \tilde L)
\cong \tilde L^{\otimes 2}$.)

But how do ${\rm Spin}^c$ structures arise at $u=1$ starting
with the underlying $SU(2)$ gauge theory on the $u$ plane?
Our next goal is to explain how
the duality transformation that leads from the variables appropriate
at large $u$ to the variables appropriate near $u=1$ can in fact
generate a ${\rm Spin}^c$ structure near $u=1$.
So far the dual object $V$ has always been a connection on a line bundle
$\tilde L$; we want to see how instead a ${\rm Spin}^c$ structure
can arise under duality.

Let us return to the starting point, and add to \hujjo\ an additional
interaction
\eqn\jubxo{{i\over 32\pi}
\sum_{\alpha}e^\alpha \int_Xd^4x\sqrt g
\epsilon^{mnpq}F_{mn}g_{\alpha\,pq}}
with integers $e^\alpha$.  The origin of this interaction will be explained
at the end of this section.

To carry out a duality transformation,
we -- as always -- introduce the two-form $G$, and replace $F$ by
$\cf= F-G$.  Then the extended Lagrangian becomes
\eqn\otopop{\eqalign{  I'= &
{i\over 16\pi}\int_Xd^4x\sqrt g \epsilon^{mnpq}W_{mn}G_{pq}\cr  &
+{i\over 16\pi}\int_Xd^4x\sqrt g\left(\bar\tau (\cf^+)^2
 -\tau (\cf^-)^2\right)+{i\over 32\pi}
\sum_{\alpha}e^\alpha \int_Xd^4x\sqrt g
\epsilon^{mnpq}\cf_{mn}g_{\alpha\,\,pq}.\cr}}

First we consider the case that $w_2(E)=c_\alpha=0$.
This will enable us to see the appearance of ${\rm Spin}^c$ structures
without extraneous complications.  After gauging $A$ to zero using
the extended gauge invariance,
the extended Lagrangian becomes
\eqn\ltopop{\eqalign{ I'=&
{i\over 16\pi}\int_Xd^4x\sqrt g \epsilon^{mnpq}W_{mn}G_{pq}
+{i\over 16\pi}\int_Xd^4x\sqrt g\left(\bar\tau (G^+)^2
 -\tau (G^-)^2\right)  \cr &
-{i\over 32\pi}
\sum_{\alpha}e^\alpha \int_Xd^4x\sqrt g
\epsilon^{mnpq}G_{mn}g_{\alpha\,pq}.\cr} }

We can eliminate the last term if we replace
$V$ by
\eqn\covvo{\tilde V= V-{1\over 2}\sum_{\alpha}e^\alpha
\theta_\alpha.}
Proceeding with the rest of the derivation, we will arrive at the same
dual Lagrangian in terms of $\tilde V$ that we have previously had
in terms of $V$.  This does not mean that the $e^\alpha$ play no material
role.  Because of the $1/2$ in \covvo, $\tilde V$ is not a connection
on a line bundle in the usual sense (unless the $e^\alpha$ are all even)
so we have obtained the same Lagrangian, but for a different kind of object.

In fact, as we will see at the end of this section, the peculiar interaction
\jubxo\ is actually present in this theory with  very special
$e^\alpha$.  To be precise,
$\sum_\alpha e^\alpha
g_\alpha/2\pi$ represents $w_2(X)$ modulo two:
\eqn\bovvo{w_2(X)=\sum_\alpha e^\alpha \left[{g_\alpha\over 2\pi}\right]
{}~{\rm mod}~ 2, }  with $[g/2\pi]$ the cohomology class of the differential
form $g/2\pi$.
Although the $1/2$ in \covvo\ means that there is not really a ``dual
line bundle'' $\tilde L$ with connection $\tilde V$, the fact that the
obstruction is $w_2(X)$ means that there is a ${\rm Spin}^c$ structure
$S_+\otimes \tilde L$ (the central part of whose curvature is $\tilde
W=d\tilde V$).
In addition, $\tilde L^{\otimes 2}$ exists as an ordinary line bundle,
so $x=-c_1(\tilde L^{\otimes 2})$ makes sense as an integral cohomology class.
Thus explaining the origin of \jubxo\ will also explain the appearance
of ${\rm Spin}^c$ structures in the dual theory near $u=1$.

\subsec{Combining The Two Effects}

So far we have studied two effects: we introduced a non-zero $w_2(E)$
by shifting the underlying Dirac quantization law as in \dico,
and we allowed for the possibility that $w_2(X)\not= 0$ by
adding a new interaction \jubxo.
It is also possible -- and natural -- to combine the two effects.
We return to the extended Lagrangian \otopop\ but use the general
Dirac quantization law of \dico.
We pick the gauge $A=\sum_\alpha c_\alpha \theta_\alpha$ and shift
$G$ by $ G\to G+\sum_\alpha c_\alpha g_\alpha$.
Instead of \ltopop\ we then get
\eqn\ultopop{\eqalign{I'=&
{i\over 16\pi}\int_Xd^4x\sqrt g \epsilon^{mnpq}W_{mn}(G_{pq}+\sum_\alpha
c_\alpha g_{\alpha\,pq})
+{i\over 16\pi}\int_Xd^4x\sqrt g\left(\bar\tau (G^+)^2
 -\tau (G^-)^2\right)  \cr &
-{i\over 32\pi}
\sum_{\alpha}e^\alpha \int_Xd^4x\sqrt g
\epsilon^{mnpq}G_{mn}g_{\alpha\,pq}.\cr} }
And we remove the last term by replacing $V$ with
\eqn\ucovvo{\tilde V= V-{1\over 2}\sum_{\alpha}e^\alpha
\theta_\alpha,}
so that the dual description involves  a ${\rm Spin}^c$ structure rather
than a line bundle.

The dependence on $w_2(E)$ can now be found as follows.  If we let
$\tilde W=d\tilde V$, then the terms in
\ultopop\ that depend on the $c_\alpha$, that is on $w_2(E)$, are
\eqn\meltdown{\Delta L =
{i\over 16\pi}\int_Xd^4x\sqrt g \epsilon^{mnpq}(\tilde
W_{mn}+ {1\over 2} \sum_\beta e^\beta g_{\beta\,mn})\cdot \sum_\alpha
c_\alpha g_{\alpha\,pq} .}
The dependence of the Donaldson invariants on $w_2(E)$ will then be
determined by a factor $e^{-\Delta L}$.
I claim that this factor coincides with the one given in \ewitten,
or in other words that
\eqn\eltdown{e^{-\Delta L}= (-1)^{x'\cdot z}}
in the notation of \ewitten, eqn. (2.17).
To justify this claim, note that the differential
form $(2\tilde W+\sum_\alpha e^\alpha g_\alpha)/2\pi$
represents the cohomology class $c_1(\tilde L^{\otimes 2})-w_2(X)$
(with a particular integral lift of
$w_2(X)$, namely $-\sum e^\alpha g_\alpha/2\pi$), so can be identified
with $-2x'$ in the notation of
\ewitten.  With also $\sum_\alpha c_\alpha g_\alpha/2\pi$
as an integral lift of $z=w_2(E)$, \eltdown\ follows from \meltdown.

So -- modulo the assumption
that on the $u$-plane there is an interaction \jubxo\ -- we have
accounted via duality for the dependence of the Donaldson invariants
on $w_2(E)$ as well as explaining the origin of ${\rm Spin}^c$ structures.
It remains to explain the origin of \jubxo.

\subsec{A Curious Minus Sign}

What remains is to explain the presence in the effective action on the
$u$-plane of the term that played a crucial role above, namely
\eqn\ippo{W={i\over 32\pi}\int_Xd^4x\sqrt g \epsilon^{mnpq} F_{mn} H_{pq},}
where $H/2\pi$ (written above as $H=\sum_\alpha e^\alpha g_\alpha/2\pi$)
represents in de Rham cohomology an integral lift of $w_2(X)$.
$W$ only depends on the isomorphism class of the line bundle $L$.

We want to study the theory with $SO(3)$ bundles $E$ of a fixed value of
$w_2(E)$.  This means that we consider only $L$ such
that $c_1(L)=w_2(E)$ modulo
two.  So picking a fixed line bundle $U$ with $c_1(U)=w_2(E)$ modulo two,
we write $L=U\otimes T^{\otimes 2}$ where now $T$ is arbitrary.

The interaction \ippo\ gives
in the path integral a factor $e^{-W}$, which is
\eqn\humbo{e^{-W}=\exp(c_1(T)\cdot w_2(X))\cdot P,}
where the factor $P$ is independent of $T$.  Our goal will
be to explain the $T$-dependent factor in \humbo, to which
the above derivations were sensitive.
The $T$-independent factors influence only the overall
(instanton-number independent)
sign convention for the Donaldson invariants.
As regards the $T$-independent factors, \ippo\ and $P$ are not the whole story
(they hardly
could be as $P$ is not necessarily real); we will find below
an additional $T$-independent interaction not
affecting anything we have said hitherto.

Since what is at issue in
\humbo\ is only a $T$-dependent minus sign, we have to be rather precise
about the treatment of fermions.
For complex fermions there are subtleties in defining the phase of the
fermion measure (these subtleties are related, among other things, to
the Adler-Bell-Jackiw anomaly and its generalizations); even for real
fermions, which we meet in the present problem, there is a subtlety with the
sign.  The subtlety arises from the fact that given orthonormal
fermi modes $\psi_1,\dots,\psi_n$,
the relation $d\psi_i\,d\psi_j=-d\psi_j\,d\psi_i$ implies that the sign
of the measure $d\psi_1\,d\psi_2\dots d\psi_n$ depends on an ordering
of the fermions up to an even permutation.

In our problem, since the instanton number of the $SU(2)$ gauge theory,
which is $k=-c_1(L)^2/4$, depends on $T$, the question of determining
the $T$-dependence of the effective action on the $u$-plane only
makes sense once one has given for all values of the instanton number
the sign of the fermion measure in the underlying $SU(2)$ theory.
For the physical, untwisted $SU(2)$ theory on ${\bf R}^4$, one usually
uses cluster decomposition to constrain the $k$-dependence of the
phase of the path integral measure.  On a four-manifold, some additional
issues arise and were analyzed by Donaldson (see \ref\dk{S. Donaldson
and P. Kronheimer, ``The Geometry Of Four-Manifolds'' (Oxford University
Press, 1990).}, p. 281) with arguments some of which will be adapted below.

I will here explain how to fix the sign of the measure in the microscopic
theory and deduce the interaction
\ippo\ in the macroscopic theory for the case that $X$ admits
an almost complex structure.  (This is a fairly mild condition,
whose import is analyzed in \dk, p. 11, and is satisfied
by all simply-connected four-manifolds with $b_2^+$ odd, a class that
includes all those with non-trivial Donaldson invariants.)
We start with the fact that the fermion fields in the untwisted
$N=2$ theory are a pair of gluinos $\alpha^i$, $i=1,2$ of positive
chirality and $U(1)_R$ charge 1, and conjugate fields $\bar\alpha_j$,
$j=1,2$ of negative chirality and charge $-1$.
($\alpha$ and $\bar\alpha$ have values in the adjoint representation of the
gauge group.)  The kinetic energy has the general form
\eqn\unbo{\int_X d^4x \sqrt g  \,\,\bar\alpha_i D\alpha^i+\dots}
where $D$ is a Dirac operator and the omitted terms involve Yukawa couplings
of a scalar field to $\alpha^2$ or $\bar\alpha^2$ (as opposed to
$\bar\alpha\cdot \alpha$ in \unbo).  As the Dirac operator is elliptic
and first order and the Yukawa terms are zeroth order, it will suffice
to define the sign of the fermion measure (or equivalently the sign of
the fermion determinant)
for the case that the scalar fields are zero and the $\alpha^2,\bar\alpha^2$
terms are absent; the dependence on the scalars then follows by continuity.
Therefore, we can ignore the mixing between $\alpha$ and $\bar\alpha$
that the Yukawa couplings would cause.

Now, to construct the ``twisted'' topological field theory, one couples
the $i$ index to the positive spin bundle $S_+$ of $X$, so that $\alpha,
\bar\alpha$ are now interpreted as spinors with values in $S_+$.
(If $X$ is not a spin manifold, the twist is needed to formulate
any theory on $X$, since on such an $X$, ordinary spinors would not exist, but
spinors with values in $S_+$ -- which can be reinterpreted as differential
forms -- still do exist.  In particular, the interpretation as differential
forms shows that the fermions have a natural real structure in the twisted
theory.)  An almost complex structure on a spin
manifold is equivalent
to a decomposition of $S_+$ as $S_+=K^{1/2}\oplus K^{-1/2}$ with
$K^{\pm 1/2}$ being line bundles, along with a choice of $K$ (or
$K^{-1}$) as the ``canonical line bundle.''  Alternatively, without assuming
that $X$ is spin, an almost complex structure is equivalent to
a ${\rm Spin}^c$ bundle (which one might informally
write as $S_+\otimes K^{1/2}$
even though the factors do not exist separately) which has a decomposition
as $K\oplus {\cal O}$, ${\cal O}$ being a trivial line bundle
and $K$ a line bundle known as the canonical line bundle of the almost
complex structure.

On an almost complex manifold, we write $\alpha=\alpha^+\oplus \alpha^-$,
with $\alpha^{\pm}$ the components of $\alpha$ in $S_+\otimes K^{\pm 1/2}$; and
similarly $\bar\alpha=\bar\alpha^+\oplus\bar\alpha^-$.
If $X$ is Kahler, the kinetic energy
\unbo\ (with Yukawa couplings suppressed) has a decomposition
\eqn\nunbo{\int_X d^4x\sqrt g \left(\bar\alpha^-D\alpha^++\bar\alpha^+
D\alpha^-\right).}
If $X$ is not Kahler, the kinetic
energy also contains $\bar\alpha^+\alpha^+$ and
$\bar\alpha^-\alpha^-$ mixing terms.  These are of zeroth order, so just
as in our dicussion
of the Yukawa terms, they can be ignored in the sense that if one defines
the sign of the fermion determinant for the Lagrangian \nunbo, then the
effect of the mixing terms can be determined by continuity.

With the various kinds of mixing terms suppressed, the eigenmodes
of $D^2$ have definite chirality and charge.  So if $\alpha^+_I$, for example,
are an orthonormal basis of eigenmodes of $\alpha^+$, we can make an
expansion
\eqn\expons{\alpha^+=\sum_I w^+_I \,\,\alpha^+_I}
with anticommuting $c$-number coefficients $w_I$.  Complex conjugation
produces eigenmodes $\alpha^-_I=\bar{\alpha^+_I}$ of $\alpha^-$, for
which we write a similar expansion:
\eqn\nexpons{\alpha^-=\sum_Iw^-_I\, \,{\alpha^-_I}.}
Now we can formally fix the sign of
the quantum integration measure for $\alpha$
by writing simply
\eqn\texpons{\prod_I \,dw^+_I \,dw^-_I.}
Similarly, one can expand
$\bar\alpha^\pm$ in orthonormal eigenmodes of $D^2$
\eqn\bexpo{\eqalign{\bar\alpha^+=\sum_J\bar w^+_J\,\,\bar\alpha^+_J\cr
                    \bar\alpha^-=\sum_J\bar w^-_J\,\,\bar\alpha^-_J\cr,}}
and take the measure for $\alpha^-$ to be formally
\eqn\texpons{\prod_I \,d\bar w^+_I \,d\bar w^-_I.}

So far we have given a prescription for the fermion measure
\eqn\hexpo{\mu=
\prod_I\,dw^+_I\, dw^-_I \,\,\prod_J\, d\bar w^+_I\, d\bar w^-_J}
which would be quite well-defined in the case of a finite set of fermi
variables -- since we have given a definite ordering
of the integration variables up to an even permutation.
With infinitely many variables, one needs the further observation
that (with the exception of the zero modes) the action of $D$ maps
modes of $\alpha$ to modes of $\bar \alpha$, preserving the eigenvalue
of $D^2$.  Therefore, except for finitely
many zero modes, the fermi modes come in groups of {\it four}, namely
modes of $\alpha^+$, $\alpha^-$, $\bar \alpha^+$, and $\bar \alpha^-$,
permuted by complex conjugation
and multiplication by $D$.  Every such group of four gives a
{\it positive} factor in the path integral, because for any real $\lambda$
the integral
\eqn\innuc{\int dw^+\,dw^-\,d\bar w^+\,d\bar w^-\,\,
\exp(\bar w^-\lambda w^++\bar w^+\lambda w^-)}
is positive.  Thus the sign of the measure on the space of zero
modes (after integrating out the non-zero modes) is simply given
by the ordering (up to an even permutation) of the zero modes in
\hexpo.

The point of this, then, is that we have made sense of the sign of  the
quantum measure for {\it all} values of the instanton number $k$,
with only a $k$-independent auxiliary choice (of almost complex structure
on $X$).  Moreover, this has been done in a way compatible with
locality and cluster decomposition, at least formally,
since the ordering of the different
types of fields given in \hexpo\ can be carried out locally.
Our choice agrees with the prescription introduced
by Donaldson, and leads to a theory that behaves well under duality
(since it will give the interaction \ippo, which has been seen to lead
to a good behavior under duality).

Now that we have defined the theory at a microscopic level,
it makes sense to reduce to slowly varying
configurations that can be described by the low energy effective action
on the $u$ plane, and to ask whether a factor \humbo\ appears in the
$T$-dependence of the effective
low energy theory.

In the low energy theory at a generic value of $u$,
the $SU(2)$ gauge group is spontaneously
broken to $U(1)$.  Correspondingly, the gluinos split up as  ``neutral''
components (valued in the Lie algebra of the unbroken $U(1)$) as
well as components of charge $\pm 1$.  The neutral components
are massless and so are included in the low energy theory; the recipe
\hexpo, restricted to those components, gives a natural (and $T$-independent)
sign of the integration
measure for the light fermions.

Of more interest are the massive fermions. They can be integrated out
to give an effective theory for the light degrees of freedom.
The effective action is complex in general (since it includes the
effective theta angle induced by the massive fermions), but the integral
over the massive fermions is real if the Higgs field is such that
the fermion mass is real.
The claim I wish to make is that if the
fermion mass is positive, the sign of the integral over the massive
fermions is precisely given in \humbo.

The massive fermions now carry {\it two} $U(1)$
charges, associated respectively
with the almost complex structure and the unbroken
gauge $U(1)$; we call these the internal and gauge charges.
So we write the expansion coefficients of the massive
fermions as $w^{\pm\,\pm}$, where the first sign refers to
the internal charge and the second to the gauge charge.
The measure defined in \hexpo\ orders
each pair of fermions with the field of positive internal charge first,
so when written out in more detail with
both charges included, it takes the form
\eqn\nexpo{\mu=\prod_Idw^{++}_I\,\,dw^{--}_I\,\prod_Jdw^{+-}_J\,dw^{-+}_J
\prod_K d\bar w^{++}_K\,d\bar w^{--}_K\,\,\prod_Ld\bar w^{+-}_L\,d\bar w
^{-+}_L.}
We could have imagined an alternative measure with the field
of positive gauge charge written first:
\eqn\unexpo{\tilde\mu=
\prod_Idw^{++}_I\,\,dw^{--}_I\,\prod_Jdw^{-+}_J\,dw^{+-}_J
\prod_K d\bar w^{++}_K\,d\bar w^{--}_K\,\,\prod_Ld\bar w^{-+}_L\,d\bar w
^{+-}_L.}
If $N_{+-}$ and $\bar N_{+-}$ are the number of modes of positive
and negative chirality, respectively, with charges $+-$, then the
relation between the two measures is
\eqn\uxxx{\mu = (-1)^{N_{+-}+\bar N_{+-}}\tilde \mu
=(-1)^{N_{+-}-\bar N_{+-}}\tilde \mu.}

Now, $N_{+-}$ and $\bar N_{+-}$ are infinite, of course,
but the difference $N_{+-}-\bar N_{+-}$  is naturally interpreted
\foot{Just as above, if one turns off the Yukawa couplings and terms
coming from non-integrability of the almost complex structure, the
action of $D$ gives a natural pairing of non-zero modes, whose effects
cancel, so only the zero modes really have to be counted.}
as the index of the Dirac operator acting on a spinor of charges $+-$,
that is, on sections of $S_+\otimes K^{1/2}\otimes L$.
Let us, as at the outset of this discussion, write $L=U\otimes T^2$
where $U$ is a fixed line bundle in the allowed class and $T$ is arbitrary.
If we denote the Dirac index for spinors with values in
$K^{1/2}\otimes L$ as ${\rm ind}(T)$, then
from the index theorem
\eqn\gurff{{\rm ind}(T)=c_1(K)\cdot c_1(T) +2c_1(T)^2+{\rm constant}.}
Hence
\eqn\umogo{\mu= {\rm constant}\cdot (-1)^{c_1(K)\cdot c_1(T)}\tilde \mu,}
with the ``constant'' being independent of $T$.

Now in fact, as we will argue momentarily, the massive fermion integral
with measure $\tilde \mu$ is positive for $u>0$.
With measure $\mu$ it therefore
has the sign given in \umogo.  The $T$-dependent part of this is a factor
$(-1)^{c_1(T)\cdot c_1(K)}$ in the effective path integral for the light
degrees of freedom.   Since, for any almost complex structure with canonical
bundle $K$, the first Chern class $c_1(K)$ reduces modulo two to
$w_2(X)$, this factor has precisely the $T$-dependence claimed in
\humbo.  This is the sought-for result.\foot{For $u>0$, the effective
theta angle as computed in \ws\ vanishes, so the sign must be attributed
to the interaction \ippo\ that appears in going to a general four-manifold.}

The argument that shows that the fermion determinant is positive
for $u$ positive
is actually closely related to a standard argument about vector-like
gauge theories such as QED.  Let us change notation slightly and refer to
the fermions of gauge charge 1 and $-1$ as $\bar \psi_\beta$ and $\psi_\beta$
as in QED ($\beta$ keeps track of labels other than gauge charge,
 such as chirality).
The measure $\tilde\mu$, with every mode
of $\bar\psi$ paired with the conjugate mode of $\psi$ (this measure
is often written formally $\tilde \mu =\prod_{x,\beta}d\bar\psi_\beta(x)
\,d\psi_\beta(x)$ or just $\tilde \mu =D\bar\psi \,D\psi$)
is the conventional measure of a vector-like
fermion.  After integrating over the non-zero modes of the Dirac operator
(which give a positive contribution because of pairing of modes of opposite
chirality) the contribution of the zero modes becomes
\eqn\imbop{\int D\bar\psi\, D\psi\,\,e^{\bar\psi M\psi},}
with everything truncated to the space of zero modes.  The integral
\imbop\ is positive if the mass matrix $M$ is positive-definite;
indeed, upon diagonalizing $M$, the integral factors as a product
of integrals
\eqn\udd{\int d\bar w \, dw\,\,\exp(m\bar w w),}
and each of these is positive if $m>0$.

Just such a situation prevails in the twisted $N=2$ model for $u>0$.
For instance,
the positive chirality fermions are vectors $\psi_m$ in the twisted
model, and the mass term is $a\,\bar\psi_m\,\psi_m$ with $a$ the Higgs field;
this is certainly positive if $a$ is positive, which leads to $u$ positive.
The negative chirality
fermions are likewise differential forms (a zero-form and a self-dual
two-form) in the twisted model, with a mass term that is similarly positive
for positive $a$.
This then implies that with measure $\tilde \mu$, the integral over
the massive fermions is positive for $u$ positive, and therefore that
with the
measure $\mu$ that arises naturally from the microscopic
theory, the integral has the sign needed in \humbo.
\bigskip

I am grateful to N. Seiberg for discussions and comments concerning many of
these matters.  I also thank R. Fintushel for some remarks on almost complex
structures.
\listrefs
\end